# Reply to: Mobility overestimation in MoS₂ transistors due to invasive voltage probes

*Hong Kuan Ng[1,2,#], Du Xiang[3,#], Ady Suwardi[1,4,#], Guangwei Hu[5], Ke Yang[6], Yunshan Zhao[7], Tao Liu[8], Zhonghan Cao[2], Huajun Liu[1], Shisheng Li[9], Jing Cao[1], Qiang Zhu[1], Zhaogang Dong[1,4], Chee Kiang Ivan Tan[1], Dongzhi Chi[1], Cheng-Wei Qiu[5], Kedar Hippalgaonkar[1,10], Goki Eda[2], Ming Yang[6,\*], Jing Wu[1,4,\**</

[1] Institute of Materials Research and Engineering, 2 Fusionopolis Way, Innovis, #08-03, Agency for Science, Technology and Research, Singapore
[2] Department of Physics, National University of Singapore, Singapore
[3] Frontier Institute of Chip and System, Fudan University, Shanghai 200438, China
[4] Department of Materials Science and Engineering, National University of Singapore, Singapore
[5] Department of Electrical and Computer Engineering, National University of Singapore, Singapore
[6] Department of Applied Physics, The Hong Kong Polytechnic University, Hong Kong
[7] NNU-SULI Thermal Energy Research Center (NSTER) & Center for Quantum Transport and Thermal Energy Science (CQTES), School of Physics and Technology, Nanjing Normal University, Nanjing, China
[8] Institute of Optoelectronics, Fudan University, Shanghai, China
[9] International Center for Young Scientists (ICYS), National Institute for Materials Science (NIMS), Tsukuba, Japan
[10] Department of Materials Science and Engineering, Nanyang Technological University, 50 Nanyang Avenue, 639798, Singapore

[#] These authors contributed equally to this work.
[*] Correspondence and requests for materials should be addressed to M. Y. (email: kevin.m.yang@polyu.edu.hk) or J. W. (email: wujing@imre.a-star.edu.sg)

In response to our recent paper[1] on the improvement of carrier mobilities in two-dimensional (2D) semiconductor transistors using rippled molybdenum disulfide (MoS$_2$), we received comments from Peng Wu (hereafter referred as PW)[2] regarding a potential "overestimation" of mobility values in our experimental method due to contact issues from partially-invasive device configuration employed in our four-probe measurement setup. The main argument revolves around whether the mobility "kink" observed in our rippled devices arises from device configuration or an intrinsic property.

In our reply, we include new experimental results and verify that the observed non-linearity in rippled-MoS$_2$ (leading to mobility kink) is an intrinsic property of a disordered system, rather than contact effects (invasive probes) or other device issues. Noting that PW's hypothesis is based on a highly ordered ideal system, transfer curves are expected to be linear, and the carrier mobility is assumed to be constant. PW's hypothesis is therefore oversimplified for disordered systems and neglects carrier-density dependent scattering physics.[3,4] Thus, it is fundamentally incompatible with our rippled-MoS$_2$, and leads to the wrong conclusion.

First, experimental results for rippled- and flat-MoS$_2$ presented in our article were obtained using the same measurement method. If the observed mobility kink was caused by contacts or device configuration issues, we would expect to observe the same phenomenon in both rippled- and flat-MoS$_2$. However, this is not the case. Additionally, we addressed the issue of contacts and demonstrated that our devices have good ohmic contact (Fig. S15 and S16 in the paper) that are consistent with literature,[5,6] thus indicating that contacts minimally affects the transistor performance.

Second, according to PW's hypothesis, if the shift in threshold voltage ($V_{th}$) were solely due to contacts, the mobility kink would be accompanied by a poorer performance and smaller source-drain current ($I_{sd}$).[7,8] However, in all rippled-MoS$_2$ devices, the $I_{sd}$ is at least an order of magnitude larger than in flat-MoS$_2$, despite similar device configuration. For instance, at room-temperature, a rippled-MoS$_2$ device (channel width ~1um, $V_{sd}$ = 0.5V) is capable of producing $I_{sd}$ reaching hundreds of microamperes, which is difficult to achieve in flat-MoS$_2$. Additionally, our results confirm no notable shift in $V_{th}$ when comparing invasive and non-invasive voltage probes, as will be discussed later. These findings proves that the observed mobility kink is indeed attributed to changes in the intrinsic property of rippled-MoS$_2$.

Third, it is evident that PW's hypothesis loses its validity at high gate voltages, indicating an oversimplification. In materials with electronic polarization and high dielectric constants, an increase in carrier density causes a softening of phonon modes, leading to enhanced electron-phonon interactions.[9] This phenomenon is primarily responsible for the observed non-linearity in rippled-MoS$_2$, and it explains the rapid decrease in mobility with increasing gate voltage. Similar non-linearities are commonly reported in 2D systems with high dielectric constant or high disorder, as extensively discussed in previous studies.[4,10–15] Take Bi$_2$O$_2$Se for example, its mobility initially peaks at ~2000 cm$^2$V$^{-1}$s$^{-1}$ before decreasing to ~200 cm$^2$ V$^{-1}$ s$^{-1}$ with increasing carrier density at room-temperature,[11] exhibiting analogous behavior to rippled-MoS$_2$. This behavior arises because mobility is carrier-density dependent. Neglecting these carrier-density dependent scattering physics in theoretical models would inevitably result in the incorrect extraction of mobility values.

The influence of contacts on carrier mobility in transition metal dichalcogenide (TMD) transistors is widely recognized and extensively investigated by the 2D community.[16,17] This is why studies on 2D transistors consistently assess and compare contact resistances with existing literature. Typically, linear I-V curves are presented to demonstrate ohmic contacts at the metal-semiconductor interface to ensure minimal impact on transistor performance. Depending on the intended application, the 2D community employs various device configurations (invasive, partially-invasive and non-invasive) for transistor measurements, each offering its own advantages and disadvantages.[11,18,19]

To provide further evidence, we followed PW's suggestion and performed additional experiments. Here, we present a direct comparison between two devices: Device 1 with a non-invasive configuration (as proposed by PW, Fig. 1a) and Device 2 with a partially-invasive configuration (as in our paper[1], Fig. 1d) made using the same MoS$_2$ flake. We first measured the 2P mobility of both devices which represent fully-invasive configuration that induces the largest disturbance to electric potential distribution within the channel. The 2P mobility values are 536 and 618 cm$^2$ V$^{-1}$ s$^{-1}$ respectively (Fig. 1c,f), consistent with our reported device variation. Next, we measured the 4P mobility for Device 1 (non-invasive) where disturbance to electric potential distribution within the channel is minimal, and for Device 2 (partially-invasive) where the disturbance to electric potential distribution is intermediate.[20] If contact effects are dominant as suggested by PW, the 4P mobility values in Device 1 (non-invasive) and Device 2 (partially-invasive) are expected to be substantially smaller than their 2P mobility values. In addition, a more significant drop in 4P mobility from its own 2P mobility is expected in Device 1 compared to Device 2 since it contains the largest disturbance to electric potential distribution within the channel. However, no notable differences between 4P and 2P mobility values in both devices are observed (Table 1). In Device 1 (non-invasive), the difference between 4P mobility (541 cm$^2$ V$^{-1}$ s$^{-1}$) and 2P mobility (536 cm$^2$ V$^{-1}$ s$^{-1}$) is only ~1% (Fig. 1c). Similarly, in Device 2 (partially-invasive), the difference between 4P mobility (596 cm$^2$ V$^{-1}$ s$^{-1}$) and 2P mobility (618 cm$^2$ V$^{-1}$ s$^{-1}$) is only ~4% (Fig. 1f). All within the reasonable variation due to the channel geometry measurement. This indicates that invasive voltage probe does not play a crucial role in determining the measured carrier mobility, as opposed to PW's suggestion. Moreover, no significant shift in $V_{th}$ between transfer curves of invasive and non-invasive configurations is observed. This is supported by the consistent mobility peak shift of 2V between 2P and 4P measurements in Device 1 (non-invasive) and Device 2 (partially-invasive) (Fig. 1b,e). It is also worth noting that the variation in mobility between the presented two devices is attributed to varying lattice distortion arising from the distribution of ripples, rather than an overestimation due to contact geometry. These findings reveal that PW's hypothesis is invalid in our rippled-MoS$_2$.

Furthermore, PW misunderstood the definition of $V_{th}$ and its distinction from the zero carrier density position in Hall measurements. Due to residual carrier density, this position might not even exist in FETs with a small bandgap material at room temperature. As such, the extracted $V_{th}$ at the zero carrier density position from Hall measurements can be easily shifted, for example, to -200V for a MoS$_2$ device that is already OFF at -10V gate voltage.[18] Hence, the $V_{th}$ estimated from Hall experiments by PW is invalid. This is corroborated by our results below where no notable shift in $V_{th}$ is observed in different device configurations, as well as the divergence of the calculated electrical conductivity at high gate voltages (>15 V) in PW's hypothesis.

In summary, the enhanced carrier mobilities showcased in our paper[1] are supported by the comparison between non-invasive and invasive device configurations, demonstrating that invasive voltage probes do not play a significant role in charge transport within our rippled devices. Thus, we stand by our interpretation that the enhanced carrier mobilities are due to suppressed phonon scattering and enhanced intrinsic dielectric constant originating from the rippled-structure-induced electronic polarization.

## Data availability

All data generated or analyzed during this study are included in the published article. Source data are provided with this paper.

## Competing interests

The authors declare no competing interests.

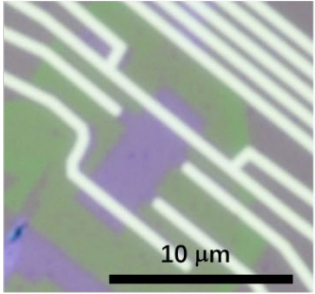
(a) Device 1: Non-invasive

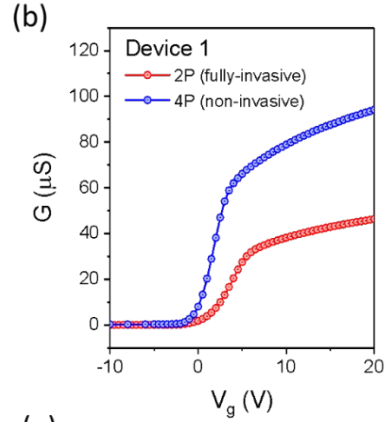
(b)

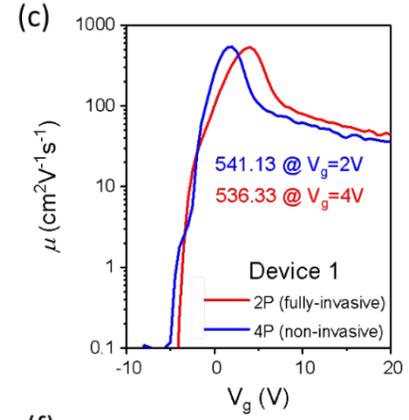
(c)

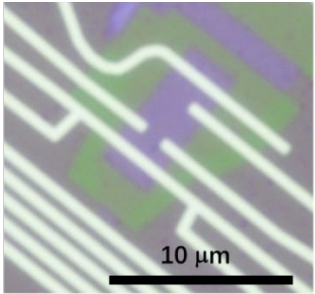
(d) Device 2: Partially-invasive

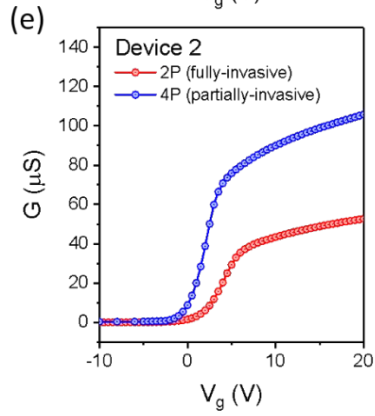
(e)

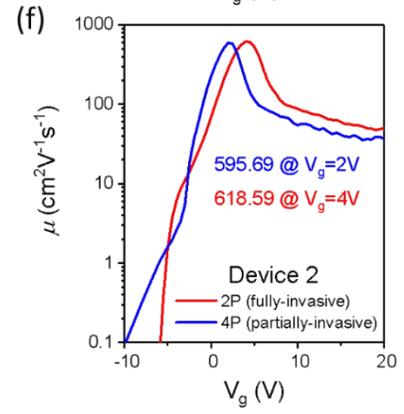
(f)

| TABLE I: Comparison of mobility | | | |
|---|---|---|---|
|  | 2P mobility (cm$^2$ V$^{-1}$ s$^{-1}$) | 4P mobility (cm$^2$ V$^{-1}$ s$^{-1}$) | Difference between 2P and 4P mobility |
| Device 1 | 536.33 (invasive) | 541.13 (non-invasive) | ~1% |
| Device 2 | 618.59 (invasive) | 595.69 (partially-invasive) | ~4% |